# Optimal Transmit Power and Antenna Selection to Achieve Energy-Efficient and Low-Complexity in 5G Massive MIMO Systems

Adeeb Salh[1,a)], Lukman Audah[1,b)], Nor Shahida Mohd Shah[2,c)], Qazwan Abdullah[1,d)], Noorsaliza Abdullah[1,e)], Jameel Mukred[1,f)], Shipun Hamzah[1,g)]

[1]Faculty of Electrical and Electronic Engineering, Universiti Tun Hussein Onn Malaysia, 86400 Parit Raja, Batu Pahat, Johor, Malaysia.
[2]Faculty of Applied Sciences and Technology, Universiti Tun Hussein Onn Malaysia, 84600 Pagoh, Muar, Johor, Malaysia

[b)] Corresponding author: hanif@uthm.edu.my
[a)]adebali@uthm.edu.my
[c)] shahida@uthm.edu.my
[d)]gazwan20062015@gmail.com
[e)]nsaliza@uthm.edu.my
[f)]jameel@uthm.edu.my
[g)]ahamzah@gmail.com

**Abstract.** This paper investigates joint antenna selection and optimal transmit power in multi-cell massive multiple-input multiple-output (MIMO) systems. The pilot interference and the number of activated transmit antenna selection plays an essential role in maximizing the energy efficiency (EE). We derived the closed form of maximal EE with complete knowledge of large-scale fading with maximum ratio transmission (MRT) while accounting for channel estimation and eliminated pilot contamination when M→∞. We investigated joint optimal antenna selection and optimal transmit power under minimized reuse of pilot sequences based on a novel iterative low-complexity algorithm for Lagrange multiplayer and Newton's methods. The two scenarios of achievable high data rate and total transmit power allocation are the critical design to the performance maximal EE. To analyze exact power consumption, we propose new power consumption for each antenna based on the transmit power amplifier and circuit power consumption. The simulation results show that maximal EE could be achieved using the iterative low-complexity algorithm based on the reasonable maximum transmit power when the noise power was less than the power received pilot. The proposed low-complexity iterative algorithm offers maximum EE by repeating a minimized pilot signal until the optimal antenna selection and transmission power are achieved.

## INTRODUCTION

Massive multiple- input – multiple output (MIMO)systems are very important when studying power trade-off pilot sequences for channel estimation to obtain better energy efficiency (EE). The EE is critical when taking into account circuit power consumption. Thus, antenna selection, optimal transmission power, and pilot reuse power are important aspects of improving EE in massive MIMO systems. These problems appear due to the increasing number of antennas in massive MIMO system due to inter-cell interference for CSI. Apart from that, a higher number of RF chains, at both BS and UE, consume more power due to the processing activities in the DAC, and power amplifier [1-5]. The EE can be achieved based on maximizing capacity channel for the pilot power sequences and available power budget at an increased number of antennas. In this case, the maximal EE requires determining the number of antennas and optimal transmit power that can be sent to every active user under impact pilot contamination.

The major challenges in massive MIMO systems come from pilot contamination, in which there are large number of pilot reuse sequences due to non-orthogonal pilot sequences among different cells. This problem requires a finite number of pilot reuse sequences, due to limited channel coherence. Massive MIMO systems are able to achieve higher EE due to active interference suppression at reasonable complexities [6-10]. EE can be achieved with maximized capacity channels for pilot power sequences and available power budgets for an increasing number of antennas. [11-15] Improved energy and spectral efficiency in a massive MIMO system under pilot contamination caused by pilot reuse sequences to estimate channel model for small-scale fading. [16-19] Investigated a joint pilot with transmit data power allocation for each user (UE) in an uplink to maximize EE. Moreover, they achieved a high data rate (DR) by using a large number of antennas to provide more active radio frequency chains (RF), which consumed circuit power. In our research, we first derived the closed form of maximal EE with complete knowledge of large-scale fading, while accounting for channel estimation, and eliminated pilot contamination. After that, we investigated joint optimal antenna selection and optimal transmit power by minimizing the reuse of pilot sequences based on a novel iterative low-complexity algorithm.

Notations: We use $\mathcal{CN}(.,.)$ for the circularly symmetric complex Gaussian distribution. $[\cdot]$ and $var(.)$ stand for expectation and variance operations, respectively. The Hermitian transpose is denoted by $(.)^M$.

## SYSTEM MODEL

We assumed that every cell consists of one base station (BS). Each BS contained many transmit antennas $N$ and UEs $K$ equipped with one antenna. The BS transmitted data signals simultaneously where $N \gg K$. We assumed that the channel reciprocities were the same in the uplink and downlink (DL), where the channel model $\mathbb{h}_{ljk} \in \mathbb{C}^{N \times 1}$ was able to assign a different antenna correlation to every channel between the UErs in the BS $l$ and $M$ in the BS $j$. In $g_{ljk} \in \mathbb{C}^N$, $N \times 1$ is the small-scale fading channel and $\mathbb{F}_{ljk} \in \mathbb{C}^{N \times N}$ accounts for the corresponding channel correlation matrix for large-scale fading (from BS $l$ to UE $k$ in cell $j$). The channel between BS $l$ and the $Kth$ UE in cell $j$ is given by $\mathbb{h}_{ljk} = \sqrt{\mathbb{F}_{ljk}}\, g_{ljk}$. We assumed that the BS was working with imperfect channel state information (CSI). The received signal $Y_{jk}$ of the $Kth$ UEs inside the $jth$ cell can be expressed as

$$Y_{jk} = \sqrt{\frac{\mathbb{B}_p P_d}{K}} \sum_{l=1, l \neq j}^{L} \sum_{i=1}^{K} \mathbb{h}_{ljk}^{H} \mathbb{v}_{lk} \mathbb{q}_{lk} + \mathbb{m}_{jk} \tag{2}$$

where $\mathbb{h}_{ljk}^{H}$ is the Hermitian transpose channel matrix and the active UEs $K$ in every cell use orthogonal pilot reuse to estimate the channel. The transmit signal vector of BS $\mho_{lk} = \mathbb{v}_{lk} \mathbb{q}_{lk} \in \mathbb{C}^N$, $g_{lk} \in \mathbb{C}^{N \times K}$ is the linear precoding matrix, $\mathbb{v}_{lk} \in \mathbb{C}^K \sim \mathcal{CN}(0, \mathbb{I}_K)$ is the data symbol transmitted from BS in cell $l$ to the UEs, $P_d$ is the DL transmit power. The channels can be evaluated with the pilot sequences sent by UEs. During the uplink training phase, $K$ UEs in the same cell transmit orthogonal pilot sequence, where $\mathbb{B}_p$ is the symbol of downlink pilot, and $\mathbb{m}_{jk} \sim \mathcal{CN}(0_{N \times 1}, I_N)$ is the received noise vector. The massive MIMO worked to acquire full channel knowledge based on the minimum mean square error (MMSE) properties used by [3],[12],[14]. The distributed channel $\mathbb{h}_{jjk} \sim \mathcal{CN}(0, \boldsymbol{\Psi}_{jk})$ can be estimated as

$$\boldsymbol{\Psi}_{jk} = \mathbb{F}_{llk} \left(\frac{\mathbb{I}_N}{\partial} + \sum_{m=1}^{L} \mathbb{F}_{lmk}\right)^{-1} \mathbb{F}_{ljk} \tag{3}$$

where $\partial$ represents the SINR during the training phase. Based on the correlated received pilot sequences and channel estimation, the interference could be reduced to accomplish a low complexity of channel estimation [7], [9], [15], [23] as in

$$\mathbb{h}_{jjk} = \mathbb{F}_{jjk} \boldsymbol{\Psi}_{jk} \sum_{l \neq j}^{K} \mathbb{h}_{ljk} + \frac{\mathbb{m}_{jk}}{\partial^{1/2}} \tag{4}$$

Estimating the channel required a limited number of pilot reuse sequences for channel coherence, so we used the received training signal during the same reused pilot sequences to adjacent cells, as in

$$\mathbb{w}_{ljk} = \sqrt{\frac{\mathbb{B}_p P_d}{K}} \sum_{j=1}^{L} \mathbb{h}_{ljk} + \mathbb{m}_{lk} \tag{5}$$

The MMSE estimate of the channel depended on $\omega_{ljk}$ and the matrix inversion, where $\mathbb{FF}^H = 1$. According to the training received signal for large-scale fading $\mathbb{F}^H \mathbb{w}_{ljk}$, the estimation channel proportional with MMSE is $\mathbb{h}_{ljk}^H / \|\mathbb{h}_{ljk}\| = \mathbb{g}^H_{ljk} \mathbb{w}_{ljk} / \|\mathbb{g}^H_{ljk} \mathbb{w}_{ljk}\|$. By substituting (4), and (5), the channel response was evaluated based on mitigated pilot contamination using a correlated channel matrix. We decomposed the received signal as

$$Y_{jk} = \sqrt{\frac{\mathbb{B}_p P_d}{K}} \mathbb{E}\{\mathbb{h}_{jjk}^H \mathbb{q}_{jk}\} \mathcal{V}_{jk} + \sqrt{\frac{\mathbb{B}_p P_d}{K}} \sum_{i=1, i \neq k}^{K} (\mathbb{h}_{jjk}^H \mathbb{q}_{jk} - \mathbb{E}\{\mathbb{h}_{jjk}^H \mathbb{q}_{jk}\}) \mathcal{V}_{jk} + \sqrt{\frac{\mathbb{B}_p P_d}{K}} \sum_{l=1, l \neq j}^{L} \sum_{i=1}^{K} \mathbb{h}_{ljk}^H \mathbb{v}_{lk} \mathbb{q}_{lk} + \mathbb{m}_{jk} \quad (6)$$

The achievable DR of the transmission from BS $l$ was obtained by considering Gaussian noise as the worst case of the UN channel, which can be written as

$$R_{tot} = Kb \, log_2 \left(1 + \frac{\mathbb{E}|DS|^2}{\mathbb{E}|UN|^2}\right) \quad (7)$$

where DS represent desired signal, $UN$ represent uncorrelated noise power. According to many studies [3-7],[18-23], pilot reuse sequences could be used in the average channel gain of $\mathbb{E}\left[\mathbb{h}_{ljk}^H \mathbb{q}_{lk}\right]$, where the pilot reuse sequences is correlated between the channel $\mathbb{h}_{ljk}^H$ and precoding $\mathbb{q}_{lk}$ in a neighboring cell. We derived the SINR at the $kth$ UEs; the DS for SINR can be expressed as

$$\mathbb{E}|DS|^2 = \frac{\mathbb{B}_p \rho_d}{K} \left|\mathbb{E}\left[\mathbb{h}_{jjk}^H \mathbb{q}_{jk}\right]\right|^2 \quad (8)$$

From the UN power can be expressed as

$$\mathbb{E}|UN|^2 = \frac{\mathbb{B}_p P_d}{K} \mathbb{E}\left\{\left|\mathbb{h}_{jjk}^H \mathbb{q}_{jk} - \mathbb{E}[\mathbb{h}_{jjk}^H \mathbb{q}_{jk}]\right|^2\right\} + \sum_{i=l}^{L} \frac{\mathbb{B}_p P_d}{K} \mathbb{E}\left\{\left|\mathbb{h}_{jjk}^H \mathbb{q}_{jk}\right|^2\right\} + \frac{\mathbb{B}_p P_d}{K} \sum_{l=1, l \neq j}^{L} \sum_{i=1}^{K} \mathbb{E}\left\{\left|\mathbb{h}_{ljk}^H \mathbb{q}_{lk} - \mathbb{E}[\mathbb{h}_{ljk}^H \mathbb{q}_{lk}]\right|^2\right\} + \Theta^2 \quad (9)$$

$$\mathbb{E}|UN|^2 = \frac{\mathbb{B}_p P_d}{K} \sum_{l=1, l \neq j}^{L} \sum_{i=1}^{K} \mathbb{E}\left[\left|\mathbb{h}_{ljk}^H \mathbb{q}_{lk}\right|^2\right] - \sum_{i=l}^{L} \frac{\mathbb{B}_p P_d}{K} \left|\mathbb{E}\left[\mathbb{h}_{jjk}^H \mathbb{q}_{jk}\right]\right|^2 + \Theta^2 \quad (10)$$

The SINR of UEs can be calculated as

$$Z_{jk}^{dl} = \frac{\frac{P_d \mathbb{B}_p}{K} \left|\mathbb{E}\left[\mathbb{h}_{jjk}^H \mathbb{q}_{jk}\right]\right|^2}{\frac{\mathbb{B}_p P_d}{K} \sum_{l=1, l \neq j}^{L} \sum_{i=1}^{K} \mathbb{E}\left[\left|\mathbb{h}_{ljk}^H \mathbb{q}_{lk}\right|^2\right] - \sum_{i=l}^{L} \frac{\mathbb{B}_p P_d}{K} \left|\mathbb{E}\left[\mathbb{h}_{jjk}^H \mathbb{q}_{jk}\right]\right|^2 + \Theta^2} \quad (11)$$

From the received signal, we estimated the regular channel and interference power in the closed form for MRT precoding, which can be expressed as

$$P_d \left|\mathbb{E}\left[\mathbb{h}_{ljk}^H \mathbb{q}_{lk}\right]\right|^2 = \frac{d}{N \, var(h_{ljk})} \left|\left\{\mathbb{E}\|\mathbb{h}_{ljk}\|^2\right\}\right|^2 \quad (12)$$

The noise variance was created from the large fading. We used the properties of the variance channels from [7], [10], [15]: $P_d \left|\mathbb{E}\left[\mathbb{h}_{ljk}^H \mathbb{q}_{lk}\right]\right|^2 = Nvar(\mathbb{h}_{ljk})$. Using (4), we estimated the row vector of channel response $\mathbb{h}_{ljk}$ between UEs in the $jth$ cell and the BS in the $lth$ cell, where $var(\mathbb{h}_{ljk}) = (P_d \mathbb{B}_p \mathbb{F}_{jjk}/(1 + P_d \mathbb{B}_p \sum_{i=1}^{L} \mathbb{F}_{ljk}))$. The channel CSI was performed for MMSE when the covariance channel interference was small and the pilot power was reasonable [24],[30]. The achievable DR in the closed form is expressed as

$$r^{mrt}{}_{jk} = Kb \, log_2(1 + \frac{\frac{P_d \mathbb{B}_p N}{K}(\frac{\mathbb{F}_{jjk}}{\frac{1}{P_d \mathbb{B}_p} + \sum_{i=1}^{L} \mathcal{F}_{ljk}})}{P_d \mathbb{B}_p \frac{N}{K}(\frac{\mathbb{F}_{ljk}}{\frac{1}{P_d \mathbb{B}_p} + \sum_{i=1}^{L} \mathbb{F}_{ljk}}) + \frac{P_d \mathbb{B}_p}{K}\sum_{l=1}^{L}\sum_{i=1}^{K}(\frac{\mathbb{F}_{jjk}}{\frac{1}{P_d \mathbb{B}_p} + \sum_{i=1}^{L} \mathbb{F}_{ljk}}) + \frac{\Theta^2}{KP_d}}) \tag{13}$$

Following the asymptotic analysis found in [7], the estimated channel accuracy was based on the received average signal power for desired signal $\mathbb{S} = (\mathbb{F}_{jjk}/(\frac{1}{P_d \mathbb{B}_p} + \sum_{i=1}^{L} \mathbb{F}_{ljk}))$. From (13), the first term in the denominator represents the inter-cell interference power for coherence channel $\phi_Q = (N/K)(\mathbb{F}_{ljk}/(\frac{1}{P_d \mathbb{B}_p} + \sum_{i=1}^{L} \mathbb{F}_{ljk}))$ that impacted the pilot with increased numbers of $N$; the second term in the denominator represents non-coherence inter-cell interference power $\phi_{nQ} = \frac{1}{K}\sum_{l=1}^{L}\sum_{i=1}^{K}(\mathbb{F}_{jjk}/(\frac{1}{P_d \mathbb{B}_p} + \sum_{i=1}^{L} \mathbb{F}_{lik}))$, which depends on the number of antennas $N$. The third term $n = (\Theta^2/KP_d)$ represents the additive Gaussian noise. From $\mathbb{s}$ in the denominator $(\frac{1}{P_d \mathbb{B}_p} + \sum_{i=1}^{L} \mathbb{F}_{ljk})$, the interference becomes limited in the DL, while $\sum_{i=1}^{L} \mathbb{F}_{ljk}$ represents the sum of the square of the propagation transmitted from BS in the $jth$ cell to every UE inside the $lth$ cells. Every BS provided the asymptotic DR with MRT.

$$r_{tot} = Kb \, log_2(1 + \frac{\frac{P_d \mathbb{B}_p N}{K} \mathbb{S}}{P_d \mathbb{B}_p \phi_Q + P_d \mathbb{B}_p \phi_{nQ} + \frac{\Theta^2}{KP_d}}) \tag{14}$$

From (14), we can simplify with $(\phi_Q - \phi_{nQ}) = \phi$. The closed form for achievable DR can be expressed as

$$r_{tot} = Kb \, log_2(1 + \frac{P_d \mathbb{B}_p N \, \mathbb{S}}{(P_d \mathbb{B}_p \phi + n)K}) \tag{15}$$

## Power Model

Due to the reduction in energy consumption, large numbers of antennas $N$ allow the use of low cost RF amplifiers that exploit antenna selection for each subcarrier, including the pilot sequences and data transmission $p_{PA} = \Theta p(N^*)$. The peak-to-average power ratio followed $\Theta = 3 \times ((N - 2(N)^{1/2} + 1)/(N - 1))$. The coherence interval for channel estimation involved a finite number of orthogonal pilot sequences and desired pilot signals that provided high data power to minimize the consumption power for antenna selection [15-18]. The transmit power for pilot and data transmission can be expressed as

$$p(N^*) = \frac{(P_d + \sum_{n=1}^{N} \mathbb{B}_p)}{K} \tag{16}$$

To analyze the exact power consumption, we proposed two main components at the BS: power consumption of the power amplifier and power consumption of the circuit power, which can be written as

$$p_{max} = \frac{\Theta}{K}(P_d + \sum_{n=1}^{N} \mathbb{B}_p) + Np_C \tag{17}$$

## Maximize EE for Transmission Antenna Selection and Transmission Power

In this process, we optimized the joint antenna selection and optimal transmit power to maximize EE under minimized pilot reuse sequences. Maximized EE based on channel condition achieved both DR and consumption power, where the fixed value of consumption power $p_c = (p_{BB} + p_{RF})$ depended on the number of RF chains $p_{RF}$ and baseband processing $P_{BB}$ of the activated antenna.

$$\Xi = \frac{Kb\, log_2\left(1+\frac{P_d\mathbb{B}_p N\, \mathbb{S}}{(P_d\mathbb{B}_p\phi+n)K}\right)}{\frac{\theta}{K}(P_d+\sum_{n=1}^{N}\mathbb{B}_p)+NP_C} \tag{18}$$

According to optimization theory [12],[19-24], [27], the joint transmit power is less than the maximum transmit power $P_{max}$. Thus, following (18), the EE is formulated for the optimization problem as follows.

$$s.t. \quad K \leq N \leq M \tag{18a}$$

$$\frac{\theta}{K}\left(P_d + \sum_{n=1}^{N}\mathbb{B}_p\right) + NP_C \leq P_{max} \tag{18b}$$

$$P_d \geq 0 \tag{18c}$$

$$\mathbb{B}_p \geq 0, \quad n = 1, 2, \ldots, N \tag{18d}$$

Constraints (18b) and (18d) are linear, while the EE is affected when transmitting $P_d$ and $\mathbb{B}_p$ based on $(P_d + \sum_{n=1}^{N}\mathbb{B}_p)$ and becomes quasi-concave with respect to $P_d$ or $\sum_{n=1}^{N}\mathbb{B}_p$. When the $P_d + \sum_{n=1}^{N}\mathbb{B}_p$ is less than $P_{max}$, the optimization problem can be solved. From (18), EE decreases with circuit power $P_c$ when $P_d$ and $\sum_{n=1}^{N}\mathbb{B}_p$ are fixed due to unavoidable interference for distributed UEs in the imperfect channel [22 - 26],[33],[37]. We used the Newton methods to solve the optimization problem (18- 18d). A general nonlinear fractional program can be expressed as

$$\max_{n\, \epsilon \mathbb{T}} \hat{\mathbf{e}}\,(N) = \frac{f_1(N)}{f_2(N)} \tag{19}$$

where $t \subseteq u$, $f_1, f_2 : t \to u$, $f_1(N) \geq 0$ and $f_2(N) \geq 0$. The fractional program in (19) accomplished concave-convex conditions where $f_1(N)$ was concave and $f_2(N)$ was convex on $t$; the maximum EE constraint can be formulated as:

$$s.t \quad \frac{f_1(N)}{f_2(N)} - \hat{\mathbf{e}} \geq 0 \tag{20}$$

We can rewrite (20) to get the solution of the optimization problem in terms of the fractional program, as in

$$\mathcal{T}(\hat{\mathbf{e}}) = \max_{N} f_1(N) - \hat{\mathbf{e}}\, f_2(N) = 0 \tag{21}$$

where $f_1$ and $f_2$ represent the concave, $\hat{\mathbf{e}}\, f_2(N)$ represents the fixed value with the concave, and $\hat{\mathbf{e}}$ is the optimum value of the objective function in (19). Following (18), the problem for joint optimal antenna selection can be written as

$$N^* = arg\, \max_{N} \frac{Kb\, log_2\left(1+\frac{P_d\mathbb{B}_p N\, \mathbb{S}}{(P_d\mathbb{B}_p\phi+n)K}\right)}{\frac{\theta}{K}(P_d+\sum_{n=1}^{N}\mathbb{B}_p)+NP_C} \tag{22}$$

We used a low-complexity algorithm to get the optimal selection antenna by updating the iteration for each UEs to find the root of $\mathcal{T}(\mathbf{e}_n)$, which can be estimated using Newton's method to solve the root of $\mathcal{T}(\mathbf{e}_n)$, as

$$\mathbf{e}_{n+1} = \hat{\mathbf{e}}_n - \frac{\mathcal{T}(\mathbf{e}_n)}{\mathcal{T}\,'(\mathbf{e}_n)} = \frac{f_1(N^*)}{f_2(N^*)} \tag{23}$$

where $N^*$ represents the optimal value for $\xi(\varepsilon_n)$; the function can also be represented as $f_1(N^*) = log_2(1 + (\rho_d \mathbb{B}_p N\mathbb{S}/(P_d\mathbb{B}_p\phi + n)K))$ and as $f_2(N^*) = \frac{\theta}{K}(P_d + \sum_{n=1}^{N}\mathbb{B}_p) + NP_C$. Our goal was to obtain low-complexity algorithm for optimal antenna using the constrained transmit power in (18b-18d). Moreover, the optimal antenna $N^*$ can be obtained using (24) when the value of $P_c$ is small, which does not affect the value of $P_c$.

$$N^* = \left[\frac{b}{P_c \; e_n \; ln \; 2} - \frac{(P_d \mathbb{B}_p \phi + n)}{\mathbb{S} P_d \mathbb{B}_p}\right] K \tag{24}$$

$e_n$ converges to optimal values when the constraint conditions can be achieved according to (18a- 18b). Maximum EE can be evaluated based on the transmit power with pilot reuse sequences $\max_{P_d} = \varepsilon \; s.t._{\mathbb{B}_p} \; P_d \geq 0$ from BS to the UE. Through the activity of UEs association with BS, the issue of power consumption between the UEs and BSs needs to be considered [31-36]. To address this problem, we relaxed these variables and replaced (18-18d) with the following

$$\mathbf{e}^* = \frac{r^*_{tot}(P_d)}{\frac{\theta}{K}(P_d + \sum_{n=1}^{N} \mathbb{B}_p) + NP_C} \tag{25}$$

The relaxed problem (25-25c) of transmission power can be expressed as

$$s.t \; \mathbb{C}_1: \quad r \geq r_{min} \tag{25a}$$

$$P_d > 0 \tag{25b}$$

$$\mathbb{C}_2: \frac{\theta}{K}\left(P_d + \sum_{n=1}^{N} \mathbb{B}_p\right) + NP_C \leq P_{max} \tag{25c}$$

where $\mathbf{e}^*$ represents maximum EE and $r_{min}$ is the minimum DR transmitted to $K$. According to (25), the exact value of $r^*_{tot}(P_d)$ could not be achieved until the relaxed problem was solved and the optimal $P_d^*$ was obtained. The convergence condition for EE can be achieved by updating the transmit power $P_d$. The relaxed problem proved that the EE was still quasi-concave when the total transmits power was used for pilot $P_d + \sum_{n=1}^{N} \mathbb{B}_p$. We applied DL transmit power allocation with a low-complexity algorithm by using the Lagrange dual decomposition [36 -45] method to solve the relaxed problem (25)-(25c) as follows:

$$L(P_d^*, \mathcal{Q}_1, \mathcal{Q}_2) = \frac{K \; b \; log_2\left(1 + \frac{P_d \mathbb{B}_p M \mathbb{S}}{(P_d \mathbb{B}_p \phi + n)K}\right)}{\partial P_d} - \left(\tilde{\mathbf{e}}\left(\frac{\theta}{K}(P_d + \mathbb{B}_p) + NP_C\right)\right) + \mathcal{Q}_1(r - r_{min}) - \mathcal{Q}_2\left(\frac{\theta}{K}(P_d + \mathbb{B}_p) - P_{max}\right) \tag{26}$$

The optimization power allocation in (26) can be solved using Karush–Kuhn–Tucker conditions to obtain the optimal value of transmission power $P_d$, according to [20]:

$$\frac{\partial L(P_d^*, \mathcal{Q}_1, \mathcal{Q}_2)}{\partial P_d} = \frac{\partial r_{tot}(P_d)}{\partial P_d} + \mathcal{Q}_1 \frac{\partial r}{\partial P_d} - \tilde{\mathbf{e}} \frac{\theta}{K}(P_d + \mathbb{B}_p) - \frac{\mathcal{Q}_2}{K} \geq 0 \tag{27}$$

where $\mathcal{Q}_1$ is the Lagrange multiplier vector corresponding to the DR constraints with element $P_d \geq 0$ and $\mathcal{Q}_2$ is the Lagrange multiplier corresponding to the transmit power constraint. From the Lagrange multiplier, the dual decomposition method can be written as

$$\min_{\mathcal{Q}_1, \mathcal{Q}_2} \max_{P_d} L(P_d^*, \mathcal{Q}_1, \mathcal{Q}_2) \geq 0 \tag{28}$$

The high channel gain should be allocated to all UE at transmit power. From (28), the optimal transmit power can be obtained based on fixed and iterative Lagrange multipliers, where $\mathcal{Q}_1, \mathcal{Q}_2 \geq 0$:

$$P_d^* = \left[\left(\frac{K(b + \mathcal{Q}_1)}{(\theta \tilde{\mathbf{e}} + \mathcal{Q}_2) \; ln \; 2} - \frac{(\mathbb{B}_p \phi + n)}{\mathbb{B}_p N \mathbb{S} K}\right) K\right]^{\blacksquare} \tag{30}$$

where $b$ represents the bandwidth of the baseband signal, $\tilde{\mathbf{e}}$ is the maximal value of (19), $\theta$ is the peak-to-average power ratio, $n$ is the additive Gaussian noise, and $\mathbb{s}$ is the received average signal power for the desired signal. More transmit power should be allocated for pilots to obtain reasonable channel estimation with large pilot reuse sequences when the number of antennas is fixed. In addition, when $\tilde{\mathbf{e}}$ converges to optimal values, the optimal transmit power $P_d^*$ can be obtained based on the constraints of power in (25a) and (25c).

## NUMERICAL RESULTS

In this section, we present the performance of the proposed channel model that accounts for imperfect channel estimation with MMSE. From Fig. 1, we used the number of UEs $K = 16$ in every cell and the number of pilot reuse sequences $\mathbb{B}_p = K$. We evaluated the number of transmit antennas versus EE. The EE increased with a low complexity of optimal antenna selection under minimal pilot reuse sequences. When transmission power was low, at $P_d =10$ dB, and when $\mathbb{B}_p = K = 8$, the proposed low-complexity algorithm provided smaller EE compared to when transmission power was $P_d=10$ dB and the number of UEs $\mathbb{B}_p = K = 16$. Consequently, EE improved based on the transmission of antenna $N$ with fixed transmission power. Furthermore, when the transmission power was reasonable, optimal antenna selection occurred, and was confirmed according to (18a) and (18b). From the Fig. 1, the maximal EE depended on how many antennas can be selected. When the number of multipath increased at number RF=100, the EE started to decrease because the switches of RF did not achieve sufficient resolution which limited transmission and led to exploding costs and power consumption. Consequently, to get more energy efficient the number of RF chain needed to be connected to only a subset of BS antennas. Only a few RF per whole at transmission needed to be highly promising in reaching a balance between the rate enhancement and power saving. The maximal EE became smaller as the number of relay antennas $N$ increased based on estimated channel realization and minimized pilot reuse sequences.

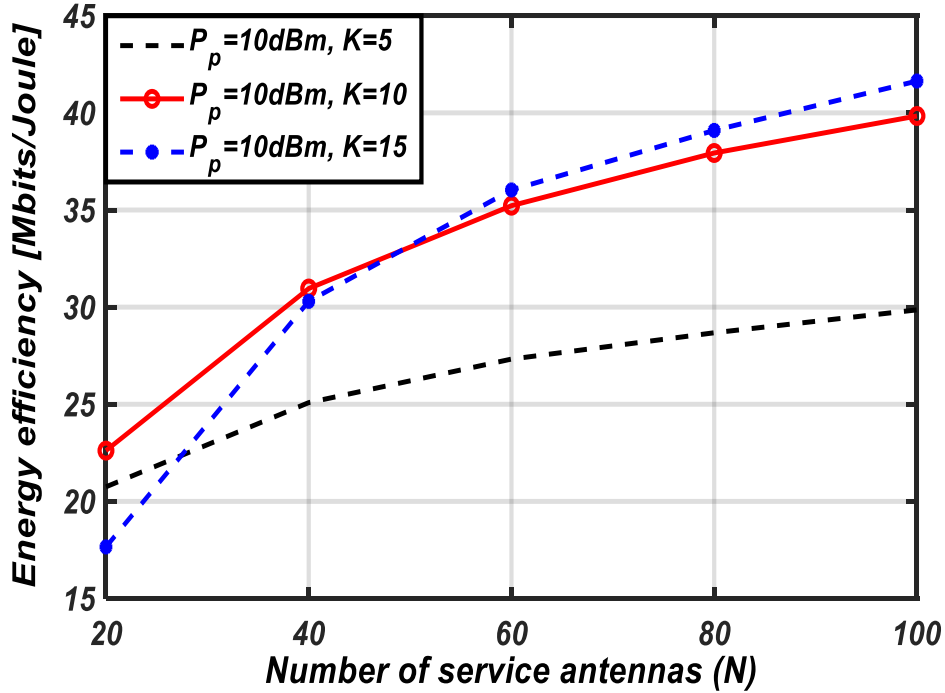

**FIGURE 1.** Energy efficiency with number of service antennas $M$ under the effect of pilot sequences user.

The proposed low-complexity algorithm resulted in better EE in the exhaustive search. Fig. 2 shows that, after obtaining a high maximum value, EE began to decrease due to an increase in transmission power and the distributed UEs inside the cells, which depended on transmission power and pilot reuse sequences power. EE was quasi-concave when the transmission power was more than the circuit power consumption and also under the optimal transmit power that satisfied the constraint in the relaxed problem (25-25c). The maximum EE can be achieved with the desired maximum transmit power and with minimized pilot reuse (where the achievable EE decreased when the pilot sequences increased).

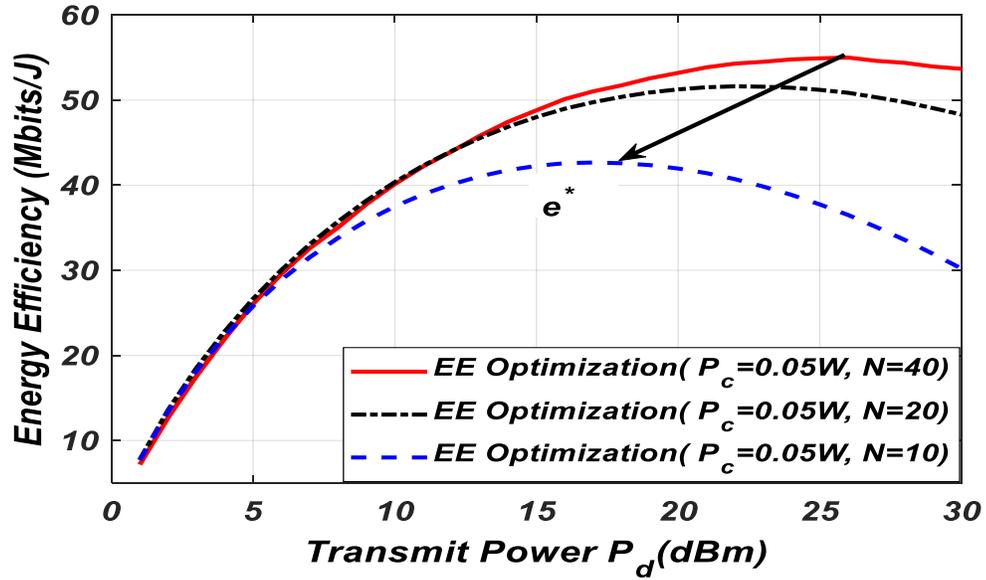

**FIGURE2**. Energy efficieny with transmission power under the effect of pilot sequences.

Alternatively, when the transmission power increased slowly, the total circuit power consumption increased, and the EE decreased, as shown in Fig.1. The EE for all curves was nonlinear to the transmit power, which required the optimal transmit power to be chosen according to (30). Alternatively, if transmission power was greater than circuit power consumption in the extra antennas, EE could be improved. Consequently, when the circuit power consumption was counterpart to the transmit power, this means that the more antennas could be used and decreased the EE. In Fig. 2 the EE started to increase and then decrease based on the circuit power consumption when compared to the transmit power and SNR. The EE started to decrease when a large number of transmit antennas was $N = 26$, and the maximum EE =53 Mb/J.

## CONCLUSION

In this paper, we presented a novel iterative low-complexity algorithm for optimal antenna selection and joint optimal transmit power under minimized pilot reuse sequences to maximize EE. Consequently, based on the number of actually used antennas, when the transmit power was more than the circuit power consumption and based on the optimize the power allocation. In this case, the EE increased monotonically and started to decrease due to high transmit power that employed more number of RF chains. Simulation results showed that the iterative low-complexity algorithm could be used to get maximize EE based on a reasonable maximum transmit power, where the noise power was less than the power received pilot; optimal antenna selection occurred when the transmission power was reasonable. EE was affected by the use of minimized pilot reuse sequences in the high SINR.

## ACKNOWLEDGMENTS

This research is supported by Universiti Tun Hussein Onn Malaysia (UTHM) under the Multisciplinary Research (MDR) Grant votH243, and part by the Ministry of Higher Education Malaysia under the Fundamental Research Grant Scheme (FRGS/1/2019/TK04/UTHM/02/8) and RMC Fund (E15501) of Universiti Tun Hussein Onn Malaysia (UTHM).